\documentstyle[aps,prl,eqsecnum,epsf]{revtex}
\begin{document}
\draft
\title{
Grain boundary motion in layered phases
}
\author{
Denis Boyer and Jorge Vi\~nals
}
\address{ 
School of Computational Science and Information Technology,\\
Florida State University, Tallahassee, Florida 32306-4120.
}
\date{\today}
\maketitle
\begin{abstract}
We study the motion of a grain boundary that separates
two sets of mutually perpendicular rolls in Rayleigh-B\'enard 
convection above onset. The problem is treated either analytically 
from the corresponding amplitude equations, or numerically by solving 
the Swift-Hohenberg equation. We find that if the rolls are curved by 
a slow transversal modulation, a net translation of the boundary follows.
We show analytically that although this motion is a nonlinear effect, 
it occurs in a time scale much shorter than that of the linear relaxation of 
the curved rolls. The total distance traveled by the boundary scales 
as $\epsilon^{-1/2}$, where $\epsilon$ is the reduced Rayleigh number.
We obtain analytical expressions for the relaxation rate of the
modulation and for the time dependent traveling velocity of the boundary,
and especially their dependence on wavenumber.
The results agree well with direct numerical solutions of the Swift-Hohenberg
equation. We finally discuss the implications of our results
on the coarsening rate of an 
ensemble of differently oriented domains in which grain boundary motion 
through curved rolls is the dominant coarsening mechanism.
\end{abstract}
 
\section{Introduction}
\label{sec:introduction}

This paper addresses the motion of a grain boundary separating two
regions of parallel convective rolls, as can be observed in a
Rayleigh-B\'enard convection cell of large aspect ratio. Each 
semi-infinite region (or grain) is comprised of roll patterns of same
wavenumber $q_0$, but the corresponding wavevectors are mutually perpendicular.
Our main focus in this paper is the relationship between local roll curvature 
and grain boundary motion. Although the model equations used in our analysis 
are appropriate for Rayleigh-B\'enard convection near onset \cite{re:cross93}, 
we expect that the qualitative features of our findings also hold in others 
systems that exhibit layered phases, such as, for example, lamellar phases in 
weakly segregated block copolymers \cite{re:seul95,re:fredrickson96}. 

Consider a disordered system brought into a layered phase, {\it e.g.} by 
a temperature quench in the case of a diblock copolymer, or by a change in 
Rayleigh number in a Rayleigh-B\'enard convection cell.
Domains comprised of rolls (in Rayleigh-B\'enard convection) or lamellae (in
block copolymers near a symmetric mixture) quickly form that have a well 
defined characteristic
wavenumber $q_{0}$ (in the case of Rayleigh-B\'enard convection near
threshold, for example, $q_{0}$ lies on the marginal stability boundary 
against a zig-zag instability \cite{re:cross84,re:cross95a}). Due to
translational and rotational invariance, the spontaneous relaxation from the
initial disordered state leads in practice to a large number of such domains,
and a sufficiently large system remains isotropic macroscopically. Such a
configuration also contains a large density of defects, such as 
grain boundaries, disclination and dislocations.

Defect motion in two dimensional layered phases has been studied extensively, 
especially in connection with the evolution of convective rolls in 
Rayleigh-B\'enard cells \cite{re:toner81,re:manneville90}.
A primary question is how defect velocities are related to
features of the background surrounding them, such as roll periodicity. 
It is well known, for example, that dislocations climb \cite{re:siggia81b}, or 
that grain boundary motion between two domains with straight rolls provides a 
wavelength selection mechanism \cite{re:manneville83b,re:tesauro87}. On the 
other hand, less attention has been paid to the relationship between defect 
motion and roll curvature \cite{re:brand84}. 
%
%
We study in this paper a particular situation that appears to be prevalent
(even if idealized) during the formation and evolution of layered structures: 
curvature induced motion of a grain boundary separating two semi-infinite 
ordered domains.
We first show that if the rolls of one of the domains are periodically 
modulated along their transverse direction, the perturbation decays in time
with a 
rate proportional to $q^{4}$, where $q$ is the wavenumber of the modulation. 
In addition, a novel feature is associated with this relaxation: the average 
position of the boundary does not remain stationary, but undergoes a net
displacement such that the size of the region with straight unperturbed rolls
increases. Analytic calculations based on the amplitude equation formalism 
show that this motion is a nonlinear effect, that it occurs in a time scale 
much shorter than the linear relaxation of the curved rolls. The total distance 
traveled by the boundary scales as $\epsilon^{-1/2}$, where $\epsilon$ is the 
reduced Rayleigh number. Consequently, the grain boundary can travel large
distances, even for very small initial perturbations.

The analysis of grain boundary motion and relaxation, while interesting in 
its own right, is also expected to contribute to our understanding of the 
coarsening of an ensemble of grains. Linear analyses of boundary or defect 
motion have already been used in the past to predict coarsening exponents 
\cite{re:elder92b,re:bray98}. The method reproduces known exponents in the 
case of models $A$ and $B$ in the lexicon of Hohenberg and Halperin 
\cite{re:hohenberg77}, and
has also been used to predict coarsening exponents for O(N) vector models
\cite{re:bray98} (see also \cite{re:bray95,re:mazenko95}). Coarsening laws for 
layered phases, on the other hand, remain largely unexplained
\cite{re:elder92,re:elder92b,re:cross95a,re:shiwa96,re:hou97,re:christensen98}.

In Section \ref{sec:model}, we describe the configuration of the grain boundary 
studied, as well as the 
model equations. Section \ref{sec:linear_stab} presents a linear stability 
analysis of long wavelength transverse modulations 
near the boundary. The numerical method used to validate our solutions is 
outlined in Section \ref{sec:numerics}, as well as the motivation for the 
nonlinear analysis given in Section \ref{sec:energy}. 
Finally, in Section \ref{sec:conclusions}, we present conclusions, and 
briefly discuss possible implications of our results on coarsening rates. 

\section{Grain boundary configuration and governing equations}
\label{sec:model}

The base or reference state is a stationary 
and planar grain boundary separating two semi-infinite domains $A$ and $B$ 
of straight convective rolls. Both sets of rolls have the same wavelength
$\lambda_0=2\pi/q_0$. It is in general expected from previous studies 
that such grain boundaries are stable against low wavenumber 
perturbations \cite{re:malomed90}.
We wish to study here the decay of a perturbation of wavenumber $q \ll q_0$ 
applied in the direction transverse to the rolls of one of the
domains, referred to as to domain $A$ (see Fig. \ref{figgb}a).
In this article, we consider for all our analytic calculations 
the particular case in which the two sets of rolls are oriented at right 
angles relative to each other in the way depicted on Fig. \ref{figgb}b.
In this case, the rolls of domain $A$ are parallel to the
grain boundary itself.
We expect that a $90^{\circ}$ grain boundary is the boundary of 
lowest energy, and hence that it is the most common in an extended system that
evolves spontaneously. However, our analysis can be generalized to cases 
such as shown on Fig. \ref{figgb}a, in which $A$ and $B$ rolls
have arbitrary angles with respect to the grain boundary, and we do not
anticipate qualitative changes in our conclusions.

Our focus is on transverse modulations since this is the simplest kind of 
perturbation that induces roll curvature without any average 
roll compression (compression is associated with longitudinal perturbations
along the roll normal). Our study differs from that of 
ref. \cite{re:malomed90} in which a grain boundary was perturbed with a 
modulation of wave vector having components both transverse and longitudinal 
to the rolls. As was shown in ref.\cite{re:malomed90}, the longitudinal 
component of the perturbation dominated the relaxation. We analyze here the 
relaxation of a pure transverse (or curvature) mode, and obtain
results that are qualitatively different from those of ref. \cite{re:malomed90}. 

Let an order parameter $\psi$ represent, for example, the vertical velocity in 
the mid-plane of 
a convecting fluid layer. Just above onset, when the reduced Rayleigh number
$\epsilon=(R-R_c)/R_c \ll 1$ ($R_{c}$ is the critical 
Rayleigh number for instability), $\psi$ can be expanded as the superposition 
of two waves with slowly varying complex amplitudes $A$ and $B$ \cite{re:cross93},
\begin{equation}
\label{psi}
\psi(x,y,t)=\frac{1}{2}\left[ A(x,y,t)\ e^{iq_0x} + B(x,y,t)\ e^{iq_0y}
+ c.c\right];\quad  A|_{x=-\infty}=0,\ B|_{x=+\infty}=0. 
\end{equation}
The amplitudes satisfy a set of two coupled Ginzburg-Landau
equations that can be derived from the equation for $\psi$ (for instance
the Swift-Hohenberg model \cite{re:swift77}) by multi scale analysis 
\cite{re:manneville83b,re:manneville90},
\begin{equation}
\label{gl}
\frac{\partial A}{\partial t} = -\frac{\delta F}{\delta \bar{A}} \quad
,\quad\quad
\frac{\partial B}{\partial t} = -\frac{\delta F}{\delta \bar{B}} ,
\end{equation}
where $\delta F/\delta \bar{A}$ is the functional derivative with respect to 
the complex conjugate of $A$, and $F$ is a Lyapunov functional,
\begin{eqnarray}\label{liap}
F & = & \int d\vec{r}\ {\cal F} = \int d\vec{r} \left\{ -\epsilon(|A|^2+|B|^2)
+\frac{g_{}}{2}(|A|^4+|B|^4)+
g_{\perp}|A|^2|B|^2+ \right. \nonumber\\
& & \left. \xi_0^2|(\partial_x-\frac{i}{2q_0}\partial^2_y)A|^2+
\xi_0^2|(\partial_y-\frac{i}{2q_0}\partial^2_x)B|^2 \right\} .
\end{eqnarray}
The coherence length $\xi_0$ is of the order of $q_0^{-1}$,
$g_{}$ and $g_{\perp}$
are interaction coefficients, and the time scale factor has been set to unity. 
These three parameters depend on the
particular model equation considered for the original field $\psi$.

The nontrivial stationary solutions $\{A_0,B_0\}$ of the governing set of 
equations (\ref{gl})-(\ref{liap}) describe a planar boundary, and depend 
only on the coordinate $x$ normal to the boundary. They are defined by,
\begin{eqnarray}
0&=&\epsilon A_0+\xi_0^2\partial_x^2 A_0-gA_0^3-g_{\perp}B_0^2A_0\label{a0}\\
0&=&\epsilon B_0-\frac{\xi_0^2}{4q_0^2}\partial_x^4B_0-gB_0^3
-g_{\perp}A_0^2B_0 \label{b0}\ .
\end{eqnarray}
This system of equations was extensively studied in refs. 
\cite{re:manneville83b,re:tesauro87}. It is a planar grain boundary of 
thickness proportional to $\xi_0/\sqrt{\epsilon}$. 
Note that the system of equations is not invariant under permutation of 
$A_0$ and $B_0$. The amplitude $A_0$ of the
rolls parallel to the interface vanishes as $\exp(x\sqrt{\epsilon}/\xi_0)$
when $x\rightarrow-\infty$, and saturates to $(\epsilon/g)^{1/2}
{\rm tanh}(x\sqrt{\epsilon}/\xi_0)$ when $x\rightarrow+\infty$. The behavior 
of the amplitude of the rolls perpendicular to the interface
is slightly different: $B_0(x)-(\epsilon/g)^{1/2}\propto
\exp(x\sqrt{\epsilon}/\xi_0)$ when $x\rightarrow-\infty$, and there is 
a point $x^*$ such that $B_0(x^*) = 0$. It is customary to define the location
of the grain boundary by the point $x^*$.
To a good approximation, $B_0(x) \simeq 0$ for $x \ge x^*$. 

We emphasize that a stationary solution only exists when the wavenumber of the
solution equals the marginal wavenumber $q_{0}$, and that at first order in 
$\epsilon$, the solutions $A_0$ and $B_0$ 
only depend on the slow spatial variable $\epsilon^{1/2}x$, but not
on the fast scale $x$. Hence, the location of the boundary and the 
phase of the rolls are independent \cite{re:manneville90}.

\section{Linear stability analysis of a 90$^{\circ}$ grain boundary}
\label{sec:linear_stab}

We linearize Eqs. (\ref{gl}) around $\{A_0,\ B_0\}$, and assume the 
following perturbed solutions (note that the form of the perturbation 
explicitly assumes that the boundary does not undergo a net average 
displacement),
\begin{eqnarray}
A(x,y,t)&=&(A_0(x)+\tilde{a})e^{i\tilde{\phi}}, \label{perta}\\
B(x,y,t)&=&B_0(x)+\tilde{b}\ . \label{pertb}
\end{eqnarray}
The phase $\tilde{\phi}$ represents a transverse distortion, as shown in 
Fig. \ref{figgb}b, and is of the general form
\begin{equation}
\tilde{\phi}=\phi(x)\cos(qy)e^{\sigma t}\ ,\quad \tilde{\phi}\ll1. 
\label{defphi}
\end{equation}
The real fields $\tilde{a}$ and $\tilde{b}$ are amplitude corrections 
and similarly read,
\begin{eqnarray}
\tilde{a}=a(x)\cos(qy)e^{\sigma t}\label{defa}\\
\tilde{b}=b(x)\cos(qy)e^{\sigma t}\label{defb} .
\end{eqnarray}
In the above expansion, we have neglected the imaginary part of $B$.
${\cal I}m(B)$ can written as $d(x)\sin(qy)e^{\sigma t}$; yet, because
of the fourth order $x$-derivative in the equation satisfied by $B$ 
(the second relation in Eq. (\ref{gl})), $d(x)$ is of order
$\epsilon$ lower than the real part function $b(x)$, and hence negligible.  
Inserting Eqs. (\ref{perta})-(\ref{pertb}) into Eqs. (\ref{gl}), 
we find,
\begin{eqnarray}
\sigma a&=&\epsilon a+\xi_0^2[\partial_x^2a-(q^4/4q_0^2)a-
(q^2/q_0)\partial_x(\phi A_0)]-(3gA_0^2+g_{\perp}B_0^2)a
-2g_{\perp}A_0B_0 b 
\label{eqa}
\\
\sigma A_0\phi&=&\xi_0^2[A_0^{-1}\partial_x(A_0^2\partial_x\phi)-
(q^4/4q_0^2)A_0\phi+(q^2/q_0)\partial_x a] \label{eqphi}\\
\sigma b&=&\epsilon b-\xi_0^2q^2b-(3gB_0^2+g_{\perp}A_0^2)b-
2g_{\perp}B_0A_0 a\ .
\label{eqb}
\end{eqnarray}
 
We further expand the solutions in power series of the small parameter $q^2$,
and define,
$a=a_0+q^2a_2+q^4a_4+..\ $; $\phi=\phi_0+q^2\phi_2+q^4\phi_4+..\ $;
$b=b_0+q^2b_2+q^4b_4+.. $. Since the limit $q\rightarrow 0$ corresponds to a
uniform translation of the interface, $\sigma=q^2\sigma_2+q^4\sigma_4+.. $. 
We now analyze the resulting equations order by order in $q^{2}$.

At order $q^0$, Eq. (\ref{eqphi}) reduces to $A_0^2\partial_x\phi_0={\rm cst}$.
Since $A_0(x)^{-2}$ diverges exponentially at $-\infty$ and 
$\partial_x\phi$ must remain finite on the whole interval $[-\infty,\infty]$,
the only admissible solution is $\partial_x\phi_0=0$. Hence
$\phi_0=$cst is a free parameter that represents the magnitude 
of the initial phase modulation. 
With the notation of Fig. \ref{figgb}b, $\phi_{0}$ is
simply related to the  magnitude of the roll deformation
$\delta x_0$ through $\phi_0 = q_0\delta x_0$ ($\delta x_0\ll\lambda_0$).
At order $q^0$, Eqs.(\ref{eqa}) and (\ref{eqb}) can be written as
$$
H_0\left(\begin{array}{l}a_0\\b_0\end{array}\right)=0.
$$
If we take the $x$-derivative of 
Eqs. (\ref{a0}) and (\ref{b0}) we recover the above equation,
with $a_0$ and $b_0$ replaced by $\partial_x A_0$ and 
$\partial_x B_0$. Hence the solutions at this order are: 
\begin{equation}
\label{ao0}
a_0(x)=\alpha_0 \partial_x A_0, \quad 
b_0(x)=\alpha_0 \partial_x B_0 ,
\end{equation}
where $\alpha_0$ is a constant to be determined from the solvability 
condition at the next order.

At order $q^2$, the solutions $a_2$ and $b_2$ satisfy,
\begin{equation}
H_0\left(\begin{array}{l}a_2\\b_2\end{array}\right)=
\left(\begin{array}{l}
(\alpha_0\sigma_2+\phi_0\xi_0^2/q_0)\partial_x A_0\\
\alpha_0(\sigma_2+\xi_0^2)\partial_x B_0
\end{array}\right)\ .
\end{equation}
The solvability condition at order $q^2$ requires that the right hand 
side vector be orthogonal to the kernel of the adjoint of $H_0$. 
Since $H_0$ is 
hermitian, and $H_0(\partial_x A_0,\partial_x B_0)=0$, we find,
\begin{equation}
\label{c1}
(\alpha_0\sigma_2+\phi_0\xi_0^2/q_0)\int_{-\infty}^{\infty}
(\partial_x A_0)^2 dx
+\alpha_0(\sigma_2+\xi_0^2)\int_{-\infty}^{\infty}(\partial_x B_0)^2 dx=0.
\end{equation}
On the other hand, at order $q^2$ one can replace $\sigma\phi$ by 
$\sigma_2 \phi_0$, and $a$ by $\alpha_0\partial_xA_0$ in Eq. (\ref{eqphi}). 
After multiplying Eq. (\ref{eqphi}) by $A_0$, and integrating over $x$, one 
finds
\begin{equation}\label{c2}
A_0^2\partial_x\phi_2=\frac{\sigma_2\phi_0}{\xi_0^2}
\int_{-\infty}^x A_0^2(u)du
-\frac{\alpha_0}{q_0}\left[A_0(x)\partial_x A_0(x)-
\int_{-\infty}^x (\partial_u A_0)^2du\right]+ C,
\end{equation}
where $C$ is a constant of integration. However,
the condition that $|\partial_x\phi_2|<\infty$ when
$x\rightarrow-\infty$ requires $C=0$. Furthermore, in the limit
$x\rightarrow+\infty$, $A_0 \rightarrow (\epsilon/g)^{1/2}$, hence 
$\int^x A_0^2(u)du \propto x$. The gradient of $\phi_2$ remains finite 
only if $\sigma_2=0.$ Therefore
Eq. (\ref{c1}) with $\sigma_2=0$ now yields the value of the solvability
constant $\alpha_{0}$,
\begin{equation}
\label{coupl}
\alpha_0=-\delta x_0\left(\int_{-\infty}^{\infty}(\partial_xA_0)^2dx\right)
\left(\int_{-\infty}^{\infty}(\partial_xB_0)^2dx\right)^{-1},
\end{equation}
where the definition $\phi_0= q_{0} \delta x_{0}$ has also been used.

To summarize our results up to this point, the perturbed amplitudes at 
order $q^0$ are given by,
\begin{eqnarray}
A & = & \left[ A_0(x)+\alpha_0\cos(qy)e^{\sigma t}\partial_x A_0(x)\right]\ e^{i\tilde{\phi}}
\simeq A_0\left(x+\alpha_0\cos(qy)e^{\sigma t}\right)
e^{iq_0\delta x_0\cos(qy)e^{\sigma t}},
\label{amp}\\
B & \simeq & B_0\left(x+\alpha_0\cos(qy)e^{\sigma t}\right), 
\label{ampb}
\end{eqnarray}
where we have used Eqs. (\ref{perta})-(\ref{pertb}) and (\ref{ao0}).
Equations (\ref{amp}) and (\ref{ampb}) show that the amplitude moduli
$|A|$ and $|B|$ of the weakly modulated rolls at this order
equal the unmodulated profiles with a simple change in the coordinate
origin. By analogy with the planar case
\cite{re:manneville83b,re:tesauro87}, we define the location of the modulated
boundary by the set of points $\{x_g,y\}$ such that $B(x_g,y)=0$. 
Eq. (\ref{ampb}) indicates that the location of the boundary is given by the 
curve
$$
x_g(y)=x^*-\alpha_0\cos(qy)e^{\sigma t}\ ,
$$
with $B_0(x^*)=0$. It is a local translation relative to the planar boundary
by $-\alpha_0\cos(qy)e^{\sigma t}$ along the $x$-axis. 
Note, on the other hand, that the lines of constant phase of $A$ are given by 
$x=cst-\delta x_0\cos(qy)e^{\sigma t}$ instead, so that
$-\delta x_0\cos(qy)e^{\sigma t}$ represents the local deformation
of the straight rolls. 
As shown by Eq. (\ref{coupl}), boundary and roll
deformations are not independent: the lengths $\alpha_0$ and 
$\delta x_0$ are related to each other through the non-uniform profiles 
$A_0$ and $B_0$. A quite non-intuitive result 
is the opposite directions of the related translations, evidenced by the 
minus sign in Eq. (\ref{coupl}). Consider for instance a point where the lines 
of constant phase of $A$ are displaced towards the region $B$; the actual
position of the boundary, however, is displaced towards
region $A$. This effect can be seen more easily on 
Figure \ref{figop} (see also Section \ref{sec:numerics}).

At order $q^4$ integration of Eq. (\ref{eqphi}) leads to, 
$$
A_0^2\partial_x\phi_4=\int_{-\infty}^x\left(\frac{\sigma_4}{\xi_0^2}
+\frac{1}{4q_0^2}\right)\phi_0A_0^2(u)du-\frac{1}{q_0}\int_{-\infty}^x
A_0(u)\partial_u a_2 du ,
$$
where the second order result $\sigma_{2} = 0$ has been used.
The second integral in the right hand side is finite, but the first one 
diverges as $x$ when
$x\rightarrow+\infty$, except if $\sigma_4/\xi_0^2+1/4q_0^2=0$. We 
therefore conclude that, at leading order in $q$
\begin{equation}
\label{sig}
\sigma=-\frac{\xi_0^2}{4q_0^2}\ q^4.
\end{equation}
This is one of the central results of this section: the
modulated boundary is expected to relax exponentially with a rate proportional
to $q^{4}$.

The linear analysis shows that the interface is always stable with respect to 
long wavelength perturbations. This is in agreement with the case studied in 
ref. \cite{re:malomed90}, although the authors concluded that the decay rate 
$\sigma \propto q^2$ instead. Such a behavior was found because the wave vector 
of the modulation considered in their work had a nonzero component in the 
longitudinal direction.
We note that the decay rate of Eq. (\ref{sig}) is in fact identical
to that of a single wave ($A$-rolls only in Figure \ref{figgb}) 
\cite{re:manneville90}. According to Eq. (\ref{sig}), the relaxation of the 
boundary is completely determined by the relaxation of the $A$ rolls far 
from it.

Unfortunately, in spite of the fact that the
main predictions of the perturbation analysis agree well with a numerical 
solution of the Swift-Hohenberg equation, the linear analysis just discussed
is not uniformly valid (we will further elaborate on this point in Section 
\ref{sec:energy}). Consider Eq. (\ref{c2}) for $\phi_2$ in the limit $x
\rightarrow+\infty$. The leading behavior is given by
\begin{equation}
\label{fphi2}
\phi_2\simeq\frac{\alpha_0}{q_0A_0^2(\infty)}\left(\int_{-\infty}^{\infty}
(\partial_u A_0)^2du\right)x\ .
\end{equation}
Therefore the phase perturbation $\phi \simeq \phi_0+q^2\phi_2$, which is
assumed to be small in the derivation of Eq. (\ref{eqphi}), diverges as $x$ 
for large $x$ through the second order correction $\phi_2$ at any finite 
time $t$. As a consequence, from Eq. (\ref{defphi}), the $y$-component of 
the wavevector of the roll pattern is,
$$
\partial_y\tilde{\phi}=-q\sin(qy)\ e^{\sigma t}(\phi_0+q^2\phi_2),
$$
and also diverges with $x$. This implies
that, far enough from the boundary, the $y$-component of the wave vector 
becomes larger than the $x$ component, given by
$q_0+\partial_x\tilde{\phi}\simeq q_0$. 
This solution represents an increasingly zig-zagging wave, which is of course 
not physical. As it is apparent from Eq. (\ref{fphi2}), the singularity arises 
because of the presence of a non-uniform amplitude 
of the pattern near the boundary $\partial_x A_0 \neq 0$. In fact, we 
think it is possible that 
the linearized system with non-constant coefficients (\ref{eqa})-({\ref{eqb}) 
does not have bounded eigenfunctions.

It appears that the failure of the linear analysis is 
related to an additional important feature of boundary motion, not taken
into account in the previous calculation: a net translation of 
the average position of the boundary towards the $A$-region, as 
shown by the numerical calculations presented in the following Section. 
After we present the numerical evidence, we
argue in Section \ref{sec:energy} that this feature cannot be
obtained at the linear level in the amplitude of the perturbation, and that
it is in fact a singular contribution in the limit $\epsilon \rightarrow 0$.

\section{Numerical solution of the Swift-Hohenberg equation}
\label{sec:numerics}

In order to test the predictions of the previous Section, and to obtain further
insight into grain boundary motion, we have numerically solved
the Swift-Hohenberg model of Rayleigh-B\'enard convection,
\cite{re:swift77} 
\begin{equation}
\label{sh}
\frac{\partial \psi}{\partial t}=
\epsilon\psi\ -\ \frac{\xi_0^2}{4q_0^2}\ (q_0^2+\nabla^2)^2\psi-\psi^3.
\end{equation}
We consider the evolution from an initial condition that corresponds to
the geometry shown in Fig. \ref{figgb}b. We note that
solving Eq. (\ref{sh}) in the case of curved rolls is much simpler than 
solving the two dimensional Ginzburg-Landau equations (\ref{gl}).

Equation (\ref{sh}) is solved numerically with a pseudo-spectral 
method. Further details on the algorithm and 
the time integration scheme can be found in ref. \cite{re:cross94b}. The
stability of the algorithm allows relatively large values of the time step, 
which is fixed to 0.4 in the dimensionless time units of Eq. (\ref{sh}). 
Equation (\ref{sh}) is then discretized on a square grid of size
$512 \times 512 $, and occasionally $256 \times 256 $. 
Spatial discretization is 
such that there are $8$ grid nodes per wavelength $\lambda_0$.
In all the following numerical examples,
we have chosen $\xi_0 = 2q_0^{-1}$, which is close to the value that
corresponds to the stress free boundary conditions in the original system of
fluid mechanical equations, from which the Swift-Hohenberg equation is derived
as an approximation.
We use periodic boundary conditions in both directions, and hence
the initial condition comprises two symmetric, well separated grain boundaries
located at $x = 1/4$ and $x = 3/4$, in units of the
system size.  We have verified that the numerical
solutions of the Swift-Hohenberg equation for the stationary straight grain
boundary coincides with those obtained by directly solving the amplitude 
equations (\ref{a0})-(\ref{b0}), with the appropriate parameters that follow
from the multi-scale analysis of the Swift-Hohenberg equation
($g=3/4$ and $g_{\perp}=3/2$).
The modulated initial condition is implemented with the help of 
the one dimensional solution, as
\begin{equation}
\psi(x,y,t=0)=A_0(x)\cos[q_0x+q_0\delta x_0\cos(qy)]+B_0(x)\cos[q_0y].
\end{equation}
 
A typical configuration is shown in Figure \ref{figop}, taken at the 
intermediate time $t=1600$, with $\epsilon=0.04$, $q=q_0/16$ and a system size 
of $256\times256$. Note that this solution
is consistent with our earlier assumption of neglecting ${\cal I}m(B)$ in the 
perturbation equations (Eq. (\ref{pertb})), 
since the $B$ rolls remain straight near the grain boundary despite of the 
transverse modulation of the $A$ rolls. 
The opposition in phase of the roll profile and of the boundary location,
given in Eq. (\ref{coupl}), is also well reproduced numerically 
Yet, this effect is easy to observe only for relatively 
large values of $\delta x_0$, as shown on Fig. \ref{figop}.

Figure \ref{figsig} shows our numerical results for the inverse
decay rate $-\sigma^{-1}$ as a function of $q$. These results are
obtained for a system of size $512\times 512$, so that the number
of rolls that separate the two grain boundaries is twice larger than in
Figure \ref{figop}. We checked that the
planar grain boundaries did not appreciably interact, and therefore
remained stationary. The value of $\sigma$ was determined
from an exponential fit to the decay of the phase of a given roll, for
a small initial $\delta x_0$. 
We first compute numerically $\sigma$ in the absence of any grain 
boundary, {\it i.e.} with the single wave $A$. We find 
very good agreement with Eq. (\ref{sig}), given that there are no adjustable 
parameters in the theoretical curve.
When grain boundaries are present, the numerical results
(square symbols) are somewhat higher, but still compare well with 
Eq. (\ref{sig}). They remain closer to the law $\sigma\propto -q^4$, 
rather than to, say, $\sigma \propto -q^5$, and certainly than to 
$\sigma\propto -q^2$.
The small discrepancy observed is probably due to the finite size of the system:
although the two grain boundaries are separated by roughly $30$ rolls, 
they slightly interact during relaxation.
As a check, we have computed $\sigma(q=3q_0/32)$ again in a 
system of size $1024\times1024$, with same value of $\lambda_0$. The difference 
between that numerical result and the theoretical curve is reduced by a 
factor of $2.5$ compared with a system size of $512\times512$.

The numerical solution reveals an additional feature which was not
included in our linear analysis: while the boundary modulation is relaxing, 
there is net motion of the grain boundary towards the $A$ roll region.
This motion is a pure amplitude mode, as the lines of 
constant roll phase remain stationary on average.
Therefore, after a long time, the modulated rolls have completely relaxed 
but the size of the region of $A$ rolls has decreased, and the grain 
boundary is located at a distance $d_{\infty}$ 
away from its initial average location. The distance $d_{\infty}$
exceeds the initial deformation $\delta x_0$, 
and it can be even significantly larger than the roll wavelength $\lambda_0$. 
In the configuration shown in Fig. \ref{figop} (system of size 
$256\times 256$), the magnitude of the perturbation was deliberately 
chosen relatively large ($\delta x_0 = \lambda_0$), so that, at late times, 
the modulated region completely disappears by annihilation of the two 
grain boundaries. 

We now report our results for $d_{\infty}$ in a bigger
system of size $512\times512$, and with $q = 3q_{0}/32$.
Figure \ref{figdinf} shows our results for the 
total traveled distance 
$d_{\infty}$ as a function of $\delta x_0$.
The complete relaxation of the perturbation requires times around $10^5$, 
and the location of the (nearly) flat boundary
is again defined to be the point $x^*$ at which $B(x^*)=0$ 
\cite{re:manneville83b}. This point if obtained from the field $\psi$
through the relation  
$B(x)=[\psi(x,y=\lambda_0)-\psi(x,y=3\lambda_0/2)]/2$ for a flat
interface (see Eq. (\ref{psi})). We observe in the figure that the total 
traveled distance increases nonlinearly with $\delta x_0$. The following
Section gives an interpretation of this feature, and further discussion 
of the results.

\section{Defect motion through energy relaxation}
\label{sec:energy}

We study in this section a more general form for the perturbed amplitudes
than those given in Eqs. (\ref{amp})-(\ref{ampb}). We assume a perturbed 
solution of
the form,
\begin{eqnarray}
A&=&A_0\left(x-\kappa\delta x(t)\cos(qy)-d(t)\right)\ 
e^{iq_0\delta x(t)\cos(qy)} \label{newa}\\
B&=&B_0\left(x-\kappa\delta x(t)\cos(qy)-d(t)\right)
\label{newb}
\end{eqnarray}
where 
\begin{equation}
\label{pert}
\delta x(t) = \delta x_0 e^{\sigma t},\quad 
\sigma = -\frac{\xi_0^2}{4q_0^2}\ q^4, \quad {\rm and}\ \kappa = 
\int(\partial_xA_0)^2 dx/ \int(\partial_xB_0)^2dx.
\end{equation} 
The quantity $d(t)$ represents the net distance traveled by the grain 
boundary at time $t$. A linear stability analysis 
based on (\ref{newa})-(\ref{newb}) along the lines of Section
\ref{sec:linear_stab} leads to $d(t)=0$ for an interface initially located 
at the origin. As we show below, the net displacement
depends nonlinearly on the initial modulation, and hence cannot 
be obtained from a linearized set of equations. We use here a different 
method that has been widely used to study of defect motion in potential
systems \cite{re:cross93}. It is based on the identity,
\begin{equation}
\label{argen}
\frac{d F}{d t}=-2\int d\vec{r} 
\left(\left|\frac{\partial A}{\partial t}\right|^2+
\left|\frac{\partial B}{\partial t}\right|^2\right)\ ,
\end{equation}
directly derived from Eqs. (\ref{gl})-(\ref{liap}).

The left hand side of Eq. (\ref{argen}) can be calculated approximately
by substituting Eqs. (\ref{newa}) and (\ref{newb}) into Eq. (\ref{liap}), 
and taking the time derivative. After some algebra, we find,
\begin{equation}
\label{lhs}
\frac{dF}{dt}=\int d\vec{r}\left\{-\dot{d}\ \partial_x{\cal F}_{0}
+\frac{\xi_0^2 q^4}{4}\cos^2(qy)\left[2\delta x\delta\dot{x}A_0^2(u)
-\delta x^2\dot{d}\ \partial_xA_0^2(u)\right]\right\},
\end{equation}
up to order $\delta x^2$. ${\cal F}_{0}$ is the free energy density of a
planar boundary, and $u = x-\kappa\delta x(t)\cos(qy)-
d(t)$. The term involving ${\cal F}_{0}$ vanishes since ${\cal F}_{0}
(x=+\infty)={\cal F}_{0}(x=-\infty)$.
In Eq. (\ref{lhs}), we have neglected the contributions from the terms
$\partial_y A_0(u)$ and $\partial_y B_0(u)$; they are of order 
$\epsilon^{1/2}$ smaller than $A_0(u)\partial_y e^{i\tilde{\phi}}$
because of the slow variations of the amplitudes compared with the roll 
periodicity. In addition, we have used the approximation $\partial_tA_0(u)
\simeq-\dot{d}\ \partial_xA_0(u)$ by
neglecting $-\kappa\delta\dot{x}(t)\cos(qy)\ \partial_xA_0(u)$: the
contributions proportional to this last term vanish
with $\int\cos(qy)dy$ at leading order in $\epsilon$ when spatial integration
over $y$ is performed in Eq. (\ref{lhs}). For the same reason, all terms
proportional to $\delta x$ in Eq. (\ref{lhs}) do not contribute.

By using similar considerations, the right hand side of Eq. (\ref{argen}) at 
leading order in $\epsilon$ and $\delta x(t)$ reads,
\begin{equation}
\label{rhs}
-2\int d\vec{r}\left\{\dot{d}^2[(\partial_xA_0(u))^2+
(\partial_xB_0(u))^2]+ (\delta \dot{x})^2 q_0^2\cos^2(qy)A_0^2(u)\right\}.
\end{equation}

By combining Eqs. (\ref{lhs}) and (\ref{rhs}), and by using 
Eq. (\ref{pert}) for $\delta x(t)$, we find that all terms that 
do not involve $\dot{d}$ cancel, and obtain
\begin{equation}
\label{v}
\dot{d}(t)=\left(\frac{\xi_0^2}{4q_0^2}q^4\right)\frac{\epsilon}{4g_{}}
\ \frac{[q_0\delta x(t)]^2}
{\int_{-\infty}^{\infty} dx\left[(\partial_x A_0)^2+
(\partial_x B_0)^2\right]}\ ,
\end{equation}
where we have used the identities $A_0^2(+\infty)=\epsilon/g$ and 
$\int_0^L dy\cos^2(qy)/L=1/2$, where $L$ is the length of the grain boundary. 
Equation (\ref{v}) shows that $\dot{d}>0$; i.e., in the configuration 
shown in Fig. \ref{figgb} the motion of the boundary is such that
the modulated rolls are progressively replaced by the advancing perpendicular 
straight rolls. Furthermore, and contrary to the usual motion 
of kinks, the velocity is not constant but decreases as 
$\delta x(t)^2$. Since at long times $\delta x(t)\simeq 0$, the boundary
eventually stops. In Appendix A, we show that Eq. (\ref{v}) generalizes the 
equation describing the
uniform motion of a grain boundary through a time independent modulated 
background. The method presented in Appendix A also provides a simpler 
alternative way to extend the result (\ref{v}) to the case in which the
transversally modulated rolls are not parallel to the grain boundary.

Equations (\ref{pert}) and (\ref{v}) can now be used to calculate the total 
distance traveled by the grain boundary $d_{\infty}$, allowing further 
comparison with the numerical results of Section \ref{sec:numerics}. We find,
\begin{equation}
\label{dinf}
d_{\infty}=\frac{\epsilon}{8g_{}}\ 
\frac{(q_0\delta x_0)^2}
{\int_{-\infty}^{\infty} dx\left[(\partial_x A_0)^2+
(\partial_x B_0)^2\right]}\ .
\end{equation}
Note that, in contrast to $\dot{d}$, $d_{\infty}$ does not depend on the
wavenumber of the modulation $q$, for small $q$.
Since $A_0(x) = \sqrt{\epsilon/g_{}}f(\sqrt{\epsilon}\ x/\xi_0)$ at
leading order in $\epsilon$, and if a similar form is assumed for $B_0$, it
is easy to see from Eqs. (\ref{v}) and (\ref{dinf}) that $\dot{d}$ and
$d_{\infty}$ are proportional to $\epsilon^{-1/2}$ \cite{fo:db1_1}. 
Hence the displacement of the boundary diverges close to onset, regardless of
the initial amplitude of the modulation.
Eqs. (\ref{v}) and (\ref{dinf}) also show that the magnitude
of this displacement depends quadratically on the perturbation $\delta x_0$,
explaining why it can not be derived from the linear analysis presented in 
Section \ref{sec:linear_stab}.  Finally, we note that the velocity 
$\dot{d}(t)$ is proportional to $ \epsilon^{-1/2} \delta x(t)^2 q^{4}$, 
at leading order in $\epsilon$, $\delta x(t)$ and $q$.

The mechanism for net grain boundary motion can be understood as the follows: 
Relaxation requires energy decrease while straight rolls have the 
same energy regardless of their orientation (both base states A and B,
are in principle degenerate). However, when the boundary is modulated
in the way depicted in
Fig. \ref{figop}, only the rolls parallel to the boundary are distorted at 
lowest order in $\epsilon$, whereas the perpendicular rolls remain straight.
Therefore, the energy of the modulated configuration can decrease by both
linear relaxation of the initially curved rolls and by nonlinear net 
displacement of the grain boundary, the effect of which is to replace curved 
rolls by straight ones. Close to onset, relaxation through the nonlinear 
mode is dominant, and even a small perturbation $\delta x_0$ 
can induce
a large boundary displacement $d_{\infty} \gg \delta x_0$.
Of course, for fixed $\epsilon$, $d_{\infty}$ goes to zero with $\delta x_0$.
It is precisely the observation that the amplitude of this nonlinear mode 
diverges at $\epsilon \rightarrow 0$ that suggests that the linearized problem 
discussed in Section \ref{sec:linear_stab} may not have any bounded 
eigenfunctions.

At fixed $\epsilon$, it is instructive to obtain the magnitude of $\delta x_0$ 
above which the velocity of the boundary is much larger than the phase velocity 
of the rolls; i.e.,
\begin{equation}
\label{adia}
\dot{d}(t=0)\gg|\delta\dot{x}(t=0)| .
\end{equation}
In this limit, boundary motion occurs at approximately
constant roll phase. Given that
$\delta \dot{x}(0)=\sigma\delta x_0$, and by using Eq. (\ref{v}) for 
$\dot{d}(0)$, Eq. (\ref{adia}) leads to 
$$
\delta x_0\gg\left(\frac{\lambda_0}{2\pi}\right)^2\frac{4g}{\epsilon}
\int_{-\infty}^{\infty} dx\left[(\partial_x A_0)^2+
(\partial_x B_0)^2\right].
$$
With the numerical solutions obtained for $A_0(x)$ and $B_0(x)$ from the 
coupled Ginzburg-Landau equations, we
find that $\delta x_0 \gg 5.0\ 10^{-2}\lambda_0$ for $\epsilon = 0.04$, 
and $\delta x_0 \gg 7.4\ 10^{-2}\lambda_0$ for $\epsilon=0.1$. Therefore 
we conclude that nonlinear motion dominates linear relaxation 
even for relatively small initial perturbations.
We recall that our perturbative calculation is valid if $\delta x_0
< \lambda_0$.

We have verified that $d_{\infty}$  does not 
appreciably depend on the wavelength of the modulation, as shown in 
Eq. (\ref{dinf}). Figure \ref{figdinf} shows our 
analytical and numerical results of the total distance traveled 
by the grain boundary as a function of $\delta x_0$. The theoretical curve has
no adjustable parameters, and is based on the numerical resolution
of Eqs. (\ref{a0})-(\ref{b0}) with the values $g=3/4$ and $g_{\perp}=3/2$ 
that correspond to the Swift-Hohenberg equation.
We performed two series of runs, for $\epsilon=0.04$ and $\epsilon=0.1$. 
At $\epsilon=0.04$, Eq. (\ref{dinf}) is in quantitative agreement
with the numerical data for small modulations. The discrepancy observed at 
large modulation indicates the breakdown of the linear relaxation of 
$\delta x(t)$: perturbations start decaying at a faster rate than 
(\ref{sig}), so that Eq. (\ref{dinf}) is over-estimating 
the traveled distance at short times. At $\epsilon = 0.1$, we expect a 
smaller displacement $d_{\infty}$, for fixed $\delta x_0$, than in the case
$\epsilon = 0.04$. This is what we observe. However, the numerical 
results show that $d_{\infty}$ does not vary continuously with $\delta x_0$, 
but rather it increases discontinuously.
As it is apparent in Figure \ref{figdinf} (crosses), 
$d_{\infty}$ takes now only multiple values of half the wavelength of the 
pattern. A lower threshold is needed to displace the grain boundary, which we
estimate to be $\delta x_c \simeq \lambda_0/7$. These observations cannot be 
explained by the present analysis, and reveal non-adiabatic effects that go
beyond the amplitude equation formalism. Such effects have been previously
observed in other studies of fronts propagation 
\cite{re:cross93,re:bensimon88b}. The underlying cause is that with 
increasing $\epsilon$ the variations of $A$ in the slow variable 
$\epsilon^{1/2}x$ do not decouple with
the fast variable describing the underlying periodic structure. In this case,
the solution for the amplitudes depends on the position of the rolls.

\section{Conclusions}
\label{sec:conclusions}

We have found an additional mechanism for grain boundary motion in layered
phases which is driven by roll curvature in the vicinity of the boundary.
We have analytically studied the motion of an isolated grain boundary 
within the amplitude equations formalism, and numerically by
direct solution of the Swift-Hohenberg equation. Transverse
perturbations of the rolls near the grain boundary decay exponentially 
in time with a 
rate proportional to $q^{4}$, where $q$ is the wavenumber of the perturbation.
In addition, we have found a nonlinear mode involving net translation of the
average location of the boundary. On the basis of functional minimization,
this mode is preferred to simple linear relaxation near onset. The net 
velocity of the grain boundary has been computed, and is given in Eq. (\ref{v}).
This result can be interpreted as giving the velocity as the ratio between a 
time dependent 
external force imposed by curved rolls, and a drag term or mobility
that depends on 
the local amplitude profile of the grain boundary. Close to onset, the drag 
term goes 
to zero as $\epsilon^{3/2}$, and makes the interface velocity (and the total 
traveled distance) diverge as $\epsilon^{-1/2}$. We have argued that this motion
cannot be explained by linear analysis.

The precise relationship between these results and the related problem of 
coarsening of layered phases requires further study. However, we can offer a few
comments based on the results presented in this paper. If
the temporal evolution of a disordered configuration were controlled by the
relaxation of an ensemble of grain boundaries moving through a background of 
curved rolls, then Eq. (\ref{v}) could be used to infer the coarsening rate. If
$v$ is a characteristic grain boundary speed, and $\kappa$ is the characteristic
curvature of the rolls ahead of the boundary, then Eq. (\ref{v}) leads to $v
\propto \kappa^{2}$. If self-similarity holds during the coarsening of the
structure, this last relationship would imply a coarsening law 
$l(t) \sim t^{1/3}$,
where $l(t)$ is any measure of the linear scale of the structure. This result
for the coarsening law disagrees with previous literature on the subject
\cite{re:elder92,re:elder92b,re:cross95a,re:shiwa96,re:hou97,re:christensen98}.

Although the configuration studied in this paper is idealized,
is is conceivable that it describes one among possibly many competing mechanisms
during coarsening. A numerical solution of the two dimensional
Swift-Hohenberg equation close to onset shows that curved rolls are essentially 
immobile due to topological constraints (mostly disclinations), 
whereas grain boundaries move over large distances. It is likely that the
motion of the latter is in part due to a background of curved
rolls with a characteristic curvature which is set by the spatial 
distribution of disclinations. Disclinations, in turn, can be
eliminated by grain boundary motion. If the characteristic length scales 
associated with both defects (grain boundary perimeter and disclination 
separation) are proportional to each other, as required by self-similarity, 
then $t^{1/3}$ 
would be a possible contribution to the coarsening law. Whether this mechanism 
would dominate the asymptotic behavior as $l \rightarrow \infty$ is being 
currently investigated.

\begin{acknowledgments}
This research has been supported by the U.S. Department of Energy, contract
No. DE-FG05-95ER14566, and also in part by the Supercomputer Computations
Research Institute, which is partially funded by the U.S.  Department of
Energy, contract No. DE-FC05-85ER25000.
\end{acknowledgments}

\appendix 
 
\section{Grain boundary motion without phase relaxation}
\label{ap:a}

Equation (\ref{v}) can be derived in an different way, which is similar to
the calculation of the velocity of a climbing dislocation \cite{re:cross93}. 
Suppose that a set of rolls is modulated along the transverse direction,
with wave vector $\vec{q}$, and is rigidly held with a time independent 
modulation amplitude $\delta x_f$. Hence,
the system is allowed to evolve by rigid translation alone, and
no roll (or boundary) relaxation is allowed. We then introduce the 
perturbations 
(\ref{newa})-(\ref{newb}) with $\delta x(t)=\delta x_f$. Equation
(\ref{argen}) can be simply recast as
\begin{equation}
\label{v1}
\dot{d}\int dy\{ {\cal F}(x=-\infty,y)-{\cal F}(x=+\infty,y)\}=
-2\dot{d}^2{\int d\vec{r}\left[|\partial_x A|^2+|\partial_x B|^2\right]}\ ,
\end{equation}
where $\dot{d}$ is the constant velocity of the grain boundary. 
The left-hand side of Eq. (\ref{v1}) represents the free energy lost per
unit time by the system as the grain boundary moves. The straight rolls, 
of lower free energy, replace curved ones. The difference $dF$
is computed that corresponds to increasing the area of the domain 
occupied by $B$ straight rolls by an amount $\dot{d}\ dt$, and by decreasing 
by the same amount the area of the domain occupied by $A$ curved rolls. 
The free energy gain is easily computed by using Eq. (\ref{liap}) with
$B(-\infty)=(\epsilon/g_{})^{1/2}$, $A(-\infty)=0$, and 
$B(+\infty)=0$, $A(+\infty)=(\epsilon/g_{})^{1/2}e^{iq_0\ 
\delta x_f\cos(qy)}$. In the right-hand side of Eq. (\ref{v1}),  
$|\partial_x A(B)|$ is replaced by $|\partial_x A_0(B_0)|$ at leading order.
Eq. (\ref{v1}) yields Eq. (\ref{v}) straightforwardly, 
where $\delta x(t)$ must be replaced by $\delta x_f$. By using
Eq. (\ref{v1}), the numerator of Eq. (\ref{v}) can now be interpreted 
as a (time dependent) external force acting on the line defect, the 
denominator being the drag term or friction coefficient. It is interesting 
to note that the more
rigorous analysis of the combined effects of roll relaxation and front 
advance that leads to Eq. (\ref{v}) gives the same expression as when 
roll relaxation is omitted. Beside being simpler,
an additional advantage of the energy argument
presented here is that it can be applied to a configuration where the set of 
rolls  $A$ and $B$ are not parallel nor perpendicular to the grain boundary. 
In this case, the external force 
$\int dy\{ {\cal F}(x=-\infty,y)-{\cal F}(x=+\infty,y)\}$ remains unchanged, by 
rotational invariance. Only the drag term would probably differ, because 
of the different grain boundary profile. However, we expect that
the scaling behavior  $\dot{d}\propto \delta x^{2} q^4$ still holds.

\bibliographystyle{prsty}

\begin{thebibliography}{10}
\bibitem{fo:db1_1}
Actually, the dependence of $\int dx(\partial_x B_0)^2$ with $\epsilon$ differs
  slightly from $\int dx(\partial_x A_0)^2\sim \epsilon^{3/2}$ (see ref.
  \protect\cite{re:manneville83b}). However, the behavior $\dot{d}\propto
  \epsilon ^{-1/2}$ remains a good approximation.
\end{thebibliography}

\begin{figure}

\begin{center}
 \caption{
    a) Grain boundary separating two set of rolls $A$ and $B$ of
same periodicity ($|\vec{q}_0|=|\vec{q}_{0}^{\ \prime}|=q_0$). The rolls of
domain $A$ are weakly curved by a transverse modulation of
wavenumber $q\ll q_0$. b) Particular case studied here, corresponding
to a $90^{\circ}$ orientation. $\delta x_0$ is the magnitude of the
phase modulation.
     }
\label{figgb}
\end{center}

\vspace{3.0cm}
\begin{center}
  
  \caption{
      Configuration with curved interfaces obtained by numerical
solution of the Swift-Hohenberg equation in a square grid with 
$256 \times 256$ nodes, with $\epsilon=0.04$, $q=q_0/16$  and
$\delta x_0=\lambda_0=8$. Because of the curved rolls, the grain 
boundaries are moving towards each other.
  }
  \label{figop}
\end{center}

\vspace{3.0cm}
\begin{center}
  \caption{
     Decay time $-\sigma^{-1}$ (in dimensionless time units) of the grain boundary
modulation as a function of wavenumber $q$. The square symbols are the numerical 
results while the solid line corresponds to Eq. (\ref{sig}). The cross symbols 
are the numerical results obtained from the modulation of a single wave, 
without any grain boundary. 
  }
  \label{figsig}
\end{center}

\newpage
\vspace{3.0cm}
\begin{center}
  \caption{
     Total distance traveled by the interface $d_{\infty}$
as a function of the magnitude
of the initial modulation $\delta x_{0}$, for $\epsilon=0.04$ and
$\epsilon=0.1$. ($q=3q_0/32$, the grid size is $512\times512$ 
and $\lambda_0=8$). The symbols are the numerical results, and the solid 
line corresponds to Eq. (\ref{dinf}).  
  }
  \label{figdinf}
\end{center}

\end{figure}

\newpage
\centerline{\epsffile{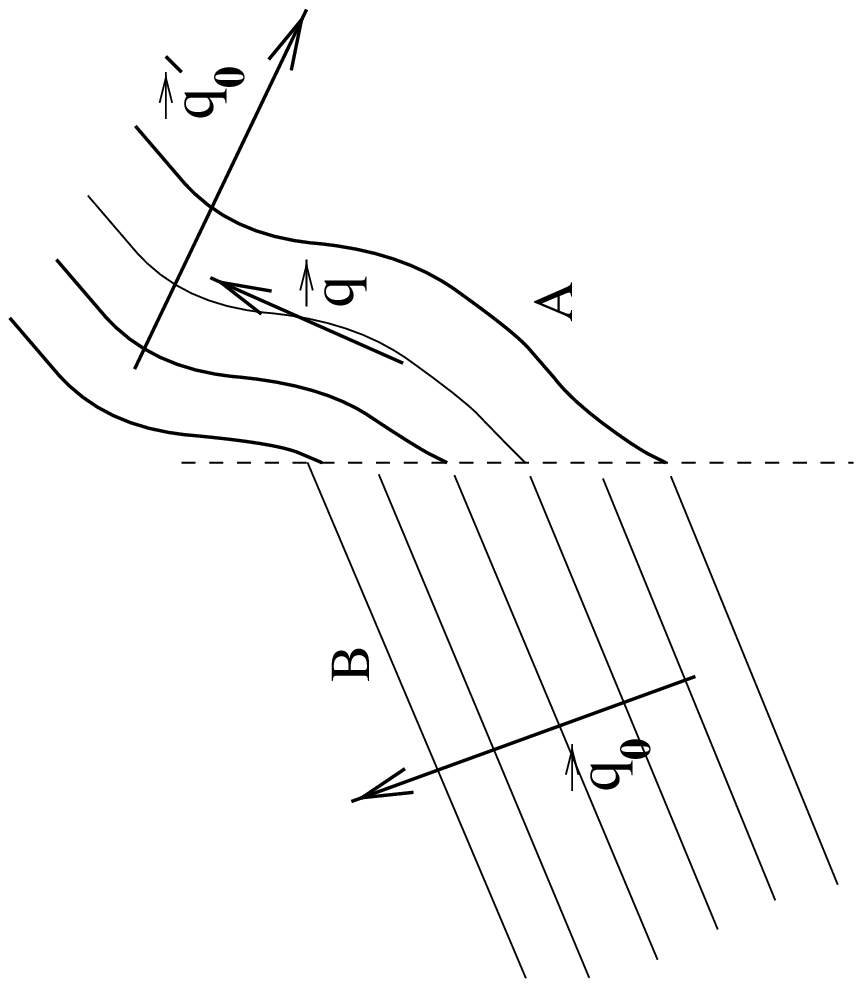}}
\centerline{Fig. 1a}

\newpage
\centerline{\epsffile{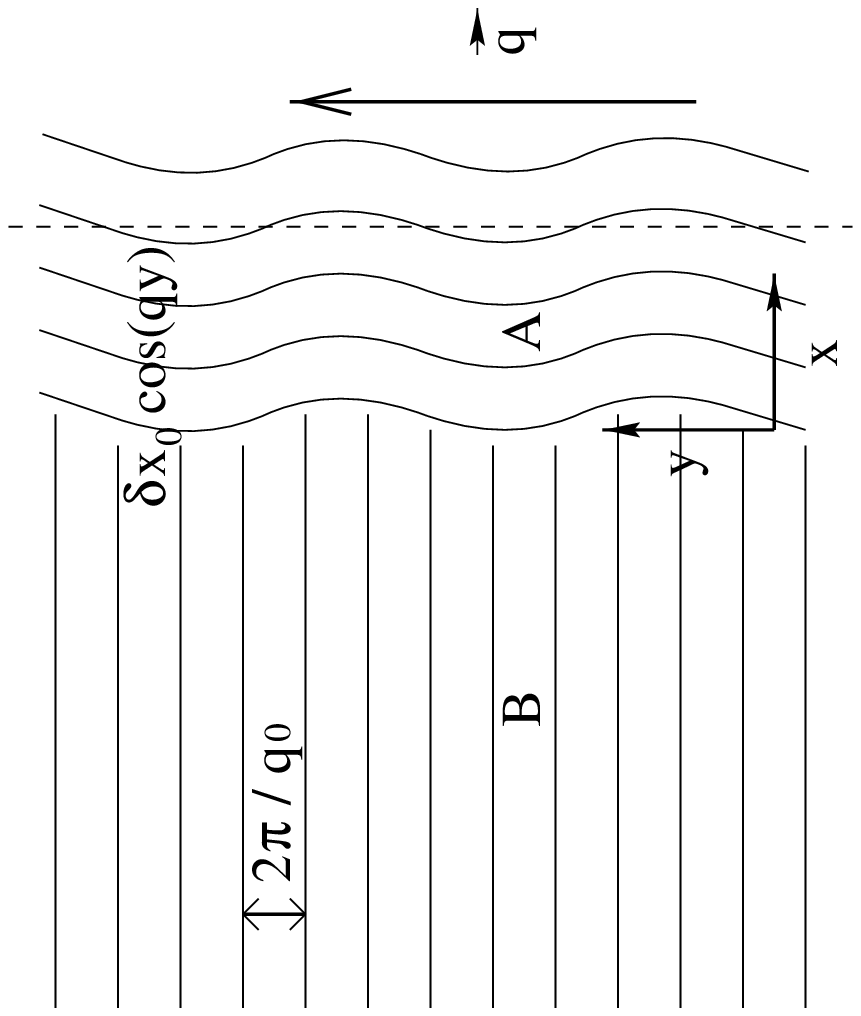}}
\centerline{Fig. 1b}

\newpage
\centerline{\epsffile{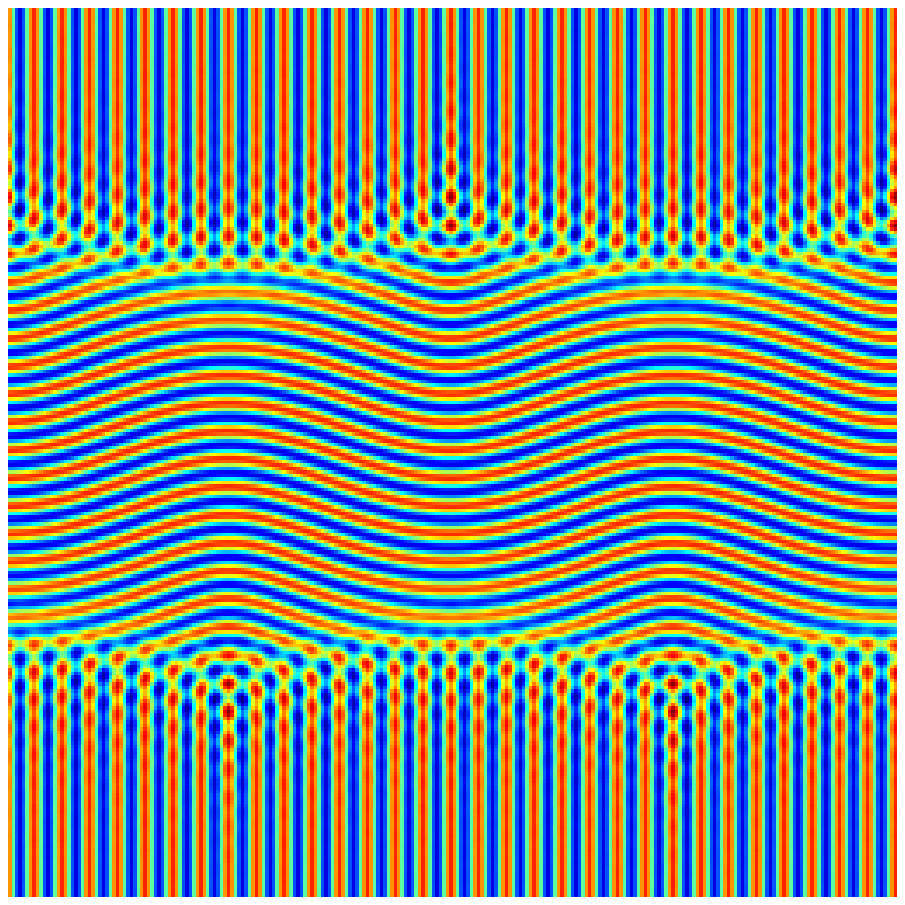}}
\centerline{Fig. 2}

\newpage
\centerline{\epsffile{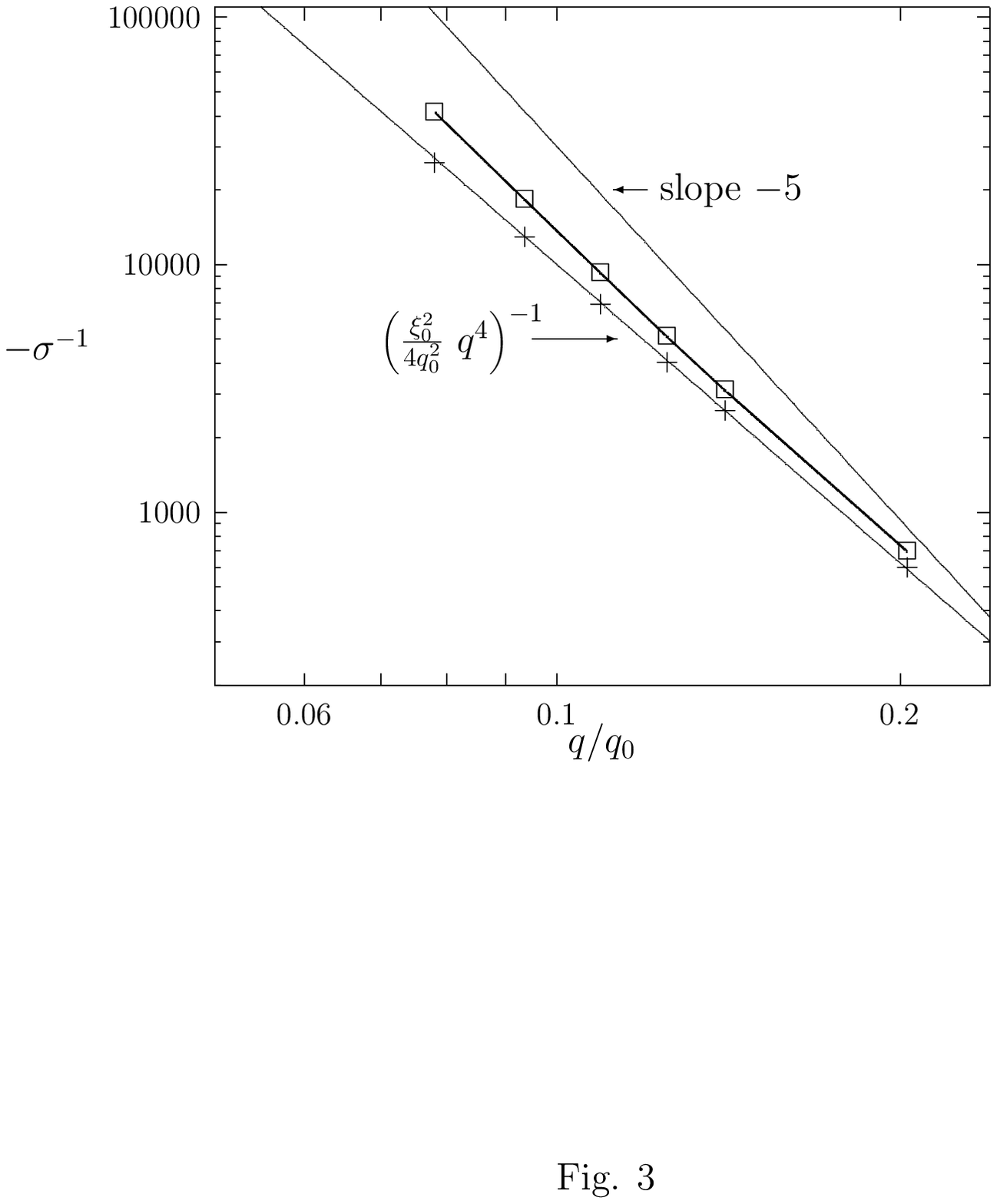}}

\newpage
\centerline{\epsffile{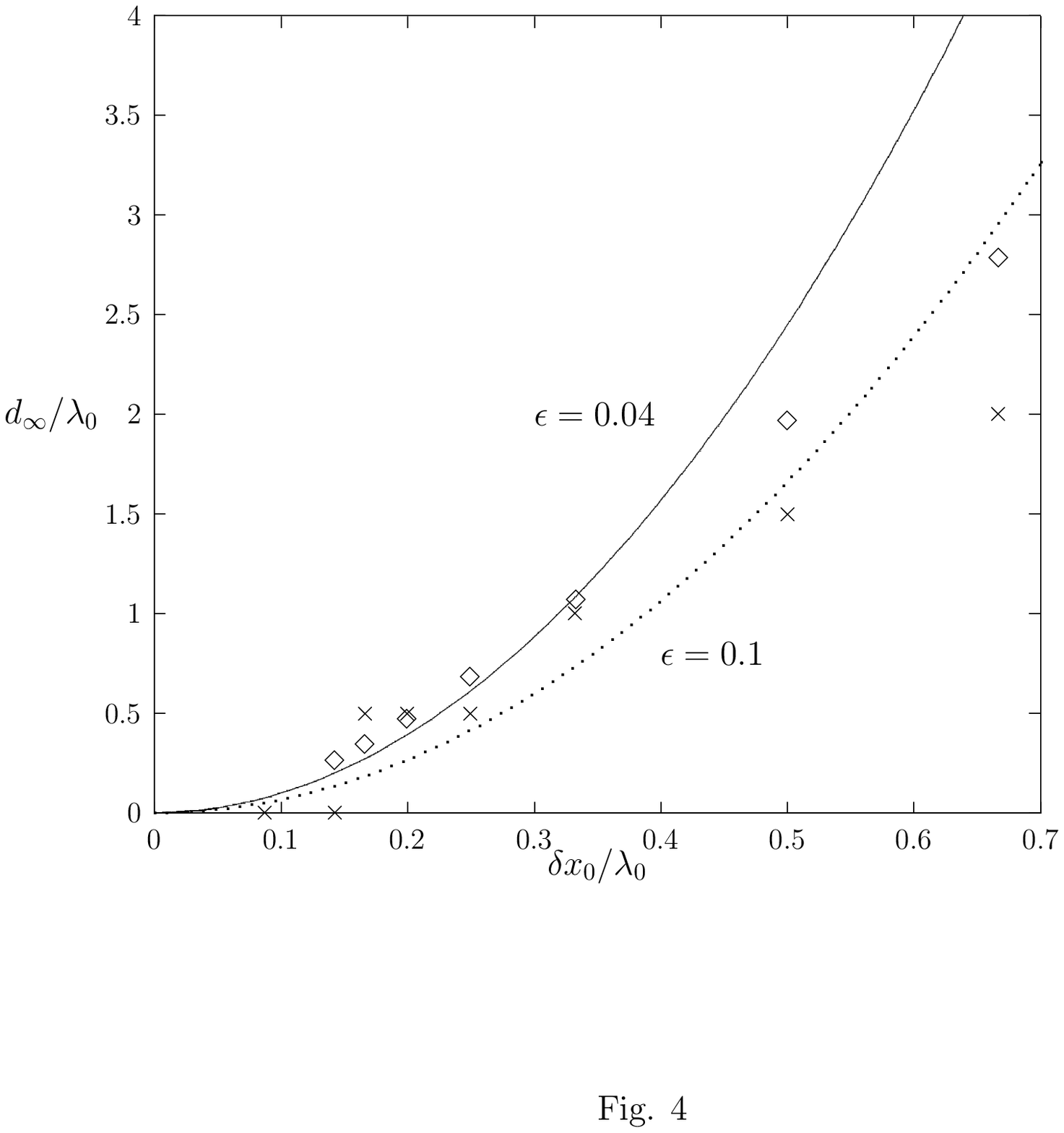}}

\end{document}